\documentclass[aps,prb,twocolumn,showpacs,letterpaper]{revtex4}
\usepackage{graphics}
\usepackage{times}

\begin{document}

\title{Coulomb Blockade Peak Spacings: 
Interplay of Spin and Dot-Lead Coupling}

\author{Serguei Vorojtsov}
\author{Harold U. Baranger}
\affiliation{Department of Physics, Duke University, 
Durham, North Carolina 27708-0305}

\date{\today}

\begin{abstract}
For Coulomb blockade peaks in the linear conductance of a quantum dot, we study the correction to the spacing between the peaks due to dot-lead coupling. This coupling can affect measurements in which Coulomb blockade phenomena are used as a tool to probe the energy level structure of quantum dots.  The electron-electron interactions in the quantum dot are described by the constant exchange and interaction (CEI) model while the single-particle properties are described by random matrix theory.  We find analytic expressions for both the average and rms mesoscopic fluctuation of the correction.  For a realistic value of the exchange interaction constant $J_{\rm s}$, the ensemble average correction to the peak spacing is two to three times smaller than that at $J_{\rm s} \!=\! 0$.  As a function of $J_{\rm s}$, the average correction to the peak spacing for an even valley decreases monotonically, nonetheless staying positive. The rms fluctuation is of the same order as the average and weakly depends on $J_{\rm s}$.  For a small fraction of quantum dots in the ensemble, therefore, the correction to the peak spacing for the even valley is negative. The correction to the spacing in the odd valleys is opposite in sign to that in the even valleys and equal in magnitude. These results are robust with respect to the choice of the random matrix ensemble or change in parameters such as charging energy, mean level spacing, or temperature.  
\end{abstract}


\maketitle

\section{Introduction}

\begin{figure}[b]
\resizebox{.45\textwidth}{!}{\includegraphics{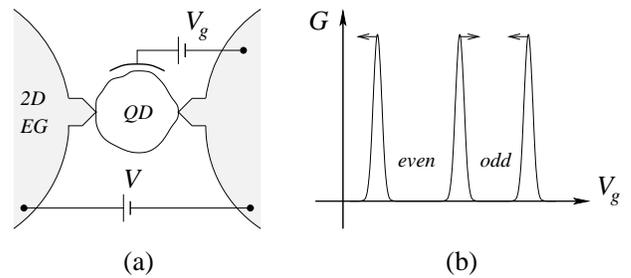}}
\caption{\label{fig:su}
(a) Scheme of the Coulomb blockade setup; (b) Oscillations of the SET linear conductance $G$ as the back-gate voltage $V_g$ is changed. ``Even'' (``odd'') corresponds to an even (odd) number of electrons in the valley.  Arrows show average shifts in the positions of the peaks' maxima due to the finite dot-lead couplings.} 
\end{figure}

Progress in nanoscale fabrication techniques has made possible not only the creation of more sophisticated devices but also greater control over their properties.  Electron systems confined to small regions -- quantum dots (QD) -- and especially their transport properties have been studied extensively for the last decade.\cite{Kou97,Del01} One of the most popular devices is a lateral quantum dot, formed by depleting the two-dimensional electron gas (2DEG) at the interface of a semiconductor heterostructure. By appropriately tuning negative potentials on the metal surface gates, one can control the QD size, the number of electrons $n$ it contains, as well as the tunnel barrier heights between the QD and the large 2DEG regions, which act as leads.  Applying bias voltage $V$ between these leads allows one to study transport properties of a single electron transistor (SET), Fig.~\ref{fig:su}(a).\cite{Kou97}

We study properties of the conductance $G$ through a QD in the linear response regime. We assume that the dot is weakly coupled to the leads: $G_{L,R} \!\ll\! e^2/h$, where $G_{L,R}$ are the conductances of the dot-lead tunnel barriers, $e \!>\! 0$ is the elementary charge, and $h$ is Planck's constant.

To tunnel onto the quantum dot, an electron in the left lead has to overcome a charging energy $E_C \!=\! e^2/2C$, where $C$ is the capacitance of the QD, a phenomenon called the Coulomb blockade.  However, if we apply voltage $V_g$ to an additional back-gate capacitively coupled to the QD, see Fig.~\ref{fig:su}(a), the Coulomb blockade can be lifted.  Indeed, by changing $V_g$ one can change the electrostatics so that energies of the quantum dot with $n$ and $n+1$ electrons become equal, and so an electron can freely jump from the left lead onto the QD and then out to the right lead.  Thus, a current event has occurred, and a peak in the conductance corresponding to that back-gate voltage, $V_{g,n+\frac{1}{2}}$, is observed.  By sweeping the back-gate voltage, a series of peaks is observed, Fig.~\ref{fig:su}(b).

In this paper we calculate the correction to the spacing between Coulomb blockade peaks due to finite dot-lead tunnel coupling.  In recent years, low-temperature Coulomb blockade experiments have been repeatedly used to probe the energy level structure of quantum dots.\cite{Kou97,Del01,Alh00}  The dot-lead tunnel coupling discussed here may influence such a measurement -- the presence of leads may change what one sees -- and so an understanding of coupling effects is needed.  One dramatic consequence is the Kondo effect in quantum dots.\cite{Gol98,Gla00} Here we assume that $T \!\gg\! T_{K}$, where $T_{K}$ is the Kondo temperature, and, therefore, do not consider Kondo physics, focusing instead on less dramatic effects that, however, survive to higher temperature.

We study an ensemble of chaotic ballistic (or chaotic disordered) quantum dots with large dimensionless conductance, $g \!\gg\! 1$.  The dimensionless conductance is defined as the ratio of the Thouless energy $E_T$ to the mean level spacing $\Delta$: $g \!\equiv\! E_T/\Delta$.\cite{Alh00}  For isolated quantum dots with large dimensionless conductance, the distribution of $g$ energy levels $\{\varepsilon_k\}$ near the Fermi level and the corresponding wave functions $\{\psi_{k}({\bf r})\}$ can be approximated by random matrix theory (RMT).\cite{Bee97,Alh00,Meh91} As will be evident from what follows, the leading contribution to the results obtained here comes from $\xi \!\equiv\! 2E_C/\Delta$ energy levels near the Fermi level; thus, if $E_{C}\lesssim E_{T}$ the statistics of these $\xi$ levels can be described by RMT.  We furthermore neglect the spin-orbit interaction and, therefore, consider only the Gaussian orthogonal (GOE) and Gaussian unitary (GUE) ensembles of random matrices.

The microscopic theory of electron-electron interactions in a quantum dot with large dimensionless conductance brings about a remarkable result.\cite{Kur00,Ale02} To leading order, the interaction Hamiltonian depends only on the squares of the following two operators: 
(i) the total electron number operator $\hat{n} = \sum c_{k\sigma}^{\dagger} c_{k\sigma}$ where $\left\{ c_{k\sigma}\right\}$ are the electron annihilation operators and $\sigma$ labels spin, and (ii) the total spin operator 
$   {\hat {\bf S}}=\frac{1}{2}\sum
      c_{k\sigma_{1}}^{\dagger}
   \left<\sigma_{1}\right|{\hat {\bf \sigma}}\left|\sigma_{2}\right>
      c_{k\sigma_{2}}
$
where $\left\{ {\hat \sigma_{i}} \right\}$ are the Pauli matrices.  The leading-order part of the Hamiltonian reads as\cite{Kur00,Ale02}
 \begin{eqnarray}
   H_{\rm int} = E_C \, \hat{n}^{2} - J_{\rm s}\, {\hat {\bf S}}^2
         \label{eq:cei}
 \end{eqnarray}
where $E_{C}$ is the redefined value of the charging energy\cite{Kur00,Usa02} and $J_{\rm s} \!>\! 0$ is the exchange interaction constant.  
Higher-order corrections are of order $\Delta/g$.\cite{Kur00,Ale02,Usa01,Usa02}
The coupling constants in (\ref{eq:cei}) are invariant with respect to different realizations of the quantum dot potential.  This ``universal'' Hamiltonian is also invariant under arbitrary rotation of the basis and, therefore, compatible with RMT.  In principal, the operator of interactions in the Cooper channel can appear in the ``universal'' Hamiltonian for the GOE case.  However, if the quantum dot is in the normal state at $T \!=\! 0$, then the corresponding coupling constant is positive and is renormalized to a very small value.\cite{Kur00,Alt85} The ``universal'' part of the Hamiltonian given by Eq.~(\ref{eq:cei}) is called the constant exchange and interaction (CEI) model.\cite{Usa01,Usa02}

The total Hamiltonian of the quantum dot in the $g \!\to\! \infty$ limit thus has two parts, the single-particle RMT Hamiltonian and the CEI model describing the interactions.  The capacitive QD-backgate coupling generates an additional term that is linear in the number of electrons:
 \begin{eqnarray}
   H_{\rm dot} = \sum_{k\sigma} \varepsilon_{k}
      \, c_{k\sigma}^{\dagger} c_{k\sigma}
         + E_C \left( \hat{n} - {\cal N} \right)^{2}
            - J_{\rm s}\, {\hat {\bf S}}^2
   \label{eq:dot}
 \end{eqnarray}
where ${\cal N} \!=\! C_{g} V_{g}/e$ is the dimensionless back-gate voltage and $C_{g}$ is the QD-backgate capacitance. 

The CEI model contains an additional exchange interaction term as compared to the conventional constant interaction model (CI model).\cite{She72,Kou97} Exchange is important as $J_{\rm s}$ is of the same order as $\Delta$, the mean single-particle level spacing. Indeed, in the realistic case of a 2DEG in a GaAs/AlGaAs heterostructure with gas parameter $r_s \!=\! 1.5$, the static random phase approximation gives $J_{\rm s}\!\approx\! 0.31\Delta$.\cite{Ore02} Therefore, as we sweep the back-gate voltage, adding electrons to the quantum dot, the conventional up-down filling sequence may be violated.\cite{Bro99,Bar00} Indeed, energy level spacings do fluctuate: If for an even number of electrons $n$ in the QD the corresponding spacing, $\varepsilon_{\frac{n}{2}+1}-\varepsilon_{\frac{n}{2}}$, is less than $2J_{\rm s}$, then it becomes energetically favorable to promote an electron to the next orbital instead of putting it in the same one; thus, a triplet state ($S \!=\! 1$) is formed.  Higher spin states are possible as well.  For $r_{s} \!=\! 1.5$ the probability of forming a higher spin ground state is $P_1(S\!>\!0)\approx 0.26$ and $P_2(S\!>\!0)\approx 0.19$ for the GOE and GUE, respectively. The lower the electron density in the QD, the larger $r_s$ and, consequently, the larger the exchange interaction constant $J_{\rm s}$.  

\begin{figure}[b]
\resizebox{.25\textwidth}{!}{\includegraphics{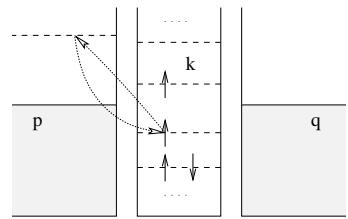}}
\caption{\label{fig:vp} 
One example of the virtual processes contributing to $E^{(2)}$.  This virtual process corresponds to an electron tunneling out of the quantum dot into the left lead and then tunneling back into the same level in the QD.} 
\end{figure}

The back-gate voltage corresponding to the conductance peak maximum ${\cal N}_{n-\frac{1}{2}}$ is found by equating the energy for $n \!-\! 1$ electrons in the dot with that for $n$ electrons:\cite{KMq}
 \begin{eqnarray}
      E_{n-1} ( {\cal N}_{n - \frac{1}{2}} ) = 
         E_{n} ( {\cal N}_{n - \frac{1}{2}} ) \;.
   \label{eq:energy}
 \end{eqnarray}
As we are interested in the effect of dot-lead coupling on these peak positions, it is natural to expand the energies perturbatively in this coupling: $E = E^{(0)} + E^{(2)} + \dots\,\,$.  One possible virtual process contributing to $E^{(2)}$ is shown in Fig.~\ref{fig:vp}.  Electron occupations of the QD ``to the left'' and ``to the right'' of the conductance peak [see Fig.~\ref{fig:su}(b)] are different; hence, the corrections $E^{(2)}$ to the energies are different. Therefore, the position of the peak maximum acquires corrections as well, ${\cal N} = {\cal N}^{(0)} + {\cal N}^{(2)} + \dots\,\,$, as does the spacing 
between two adjacent peaks.

This physical scenario has been considered by Kaminski and Glazman with the interactions treated in the CI model, i.e. neglecting exchange.\cite{Kam00} The ensemble-averaged change in the spacing and its rms due to mesoscopic fluctuations were calculated.  On average, ``even'' spacings (that is, spacings corresponding to an even number of electrons in the valley) increase, while ``odd'' spacings decrease (by the same amount):\cite{Kam00}
 \begin{eqnarray}
   2E_C~\overline{{\cal U}_{n}^{(2)}(J_{\rm s} \!=\! 0)} =
      \Delta\,\frac{g_L+g_R}{2\pi^2}\,\ln\frac{2E_C}{T},
 \end{eqnarray}
where ${\cal U}$ is the dimensionless spacing normalized by $2E_C$ and $g_{L,R} \!=\! G_{L,R}/(2e^2/h)$ are the dimensionless dot-lead conductances.

In this paper we calculate the same quantities, but with the electron-electron interactions in the QD described by the more realistic CEI model.  We find that the average change in the spacing between conductance peaks is significantly less than that predicted by the CI model.  However, the fluctuations are of the same order.  In contrast to the CI result,\cite{Kam00} for large enough $J_{\rm s}$, we find that ``even'' spacings do not necessarily increase (likewise, ``odd'' spacings do not necessarily decrease).

The paper is organized as follows.  In Sec.~\ref{sec:theham} we write down the total Hamiltonian of the system and find the condition for the tunneling Hamiltonian to be considered as a perturbation.  In Sec.~\ref{sec:approach} we describe the approach and make symmetry remarks.  In Sec.~\ref{sec:case} we perform a detailed calculation of the correction to the spacing between Coulomb blockade peaks for the $\frac{1}{2} \to 1 \to \frac{1}{2}$ spin sequence.  In Sec.~\ref{sec:averaged} we find the ensemble-averaged correction to the peak spacing.  The rms of the fluctuations of the correction to the peak spacing is calculated in Sec.~\ref{sec:fluctuations}.  In Sec.~\ref{sec:conclusions} we summarize our findings and discuss their relevance to the available experimental data.\cite{Mau99,Jeo}

\section{The Hamiltonian}
\label{sec:theham}

The Hamiltonian of the system in Fig.~\ref{fig:su}(a) consists of the QD Hamiltonian [Eq.~(\ref{eq:dot})], the leads Hamiltonian, and the tunneling Hamiltonian accounting for the dot-lead coupling:
 \begin{eqnarray}
   H = H_{\rm dot} + H_{\rm leads} + H_{\rm tun}.
      \label{eq:hamilt}
 \end{eqnarray}
The leads Hamiltonian can be written as follows:
 \begin{eqnarray}
   H_{\rm leads} = \sum_{{\bf p}\sigma}\varepsilon_{{\bf p}}
      c_{{\bf p}\sigma}^{\dagger} c_{{\bf p}\sigma} +
   \sum_{{\bf q}\sigma} \varepsilon_{{\bf q}}
      c_{{\bf q}\sigma}^{\dagger} c_{{\bf q}\sigma},
 \end{eqnarray}
where $\{\varepsilon_{\bf p}\}$ and $\{\varepsilon_{\bf q}\}$ are the one-particle energies in the left and right leads, respectively, measured with respect to the chemical potential (see Fig.~\ref{fig:vp}). We assume that the leads are large; therefore we (i) neglect electron-electron interactions in the leads and (ii) assume a continuum of states in each lead.  The tunneling Hamiltonian is\cite{Coh62}
 \begin{equation}
   H_{\rm tun} = \sum_{k{\bf p}\sigma}(t_{k{\bf p}} c_{k\sigma}^{\dagger}
         c_{{\bf p}\sigma} + {\rm h.c.}) +
      \sum_{k{\bf q}\sigma}(t_{k{\bf q}} c_{k\sigma}^{\dagger}
         c_{{\bf q}\sigma} + {\rm h.c.}),
   \label{eq:tunnelingh}
 \end{equation}
where $\{ t_{k{\bf p}}\}$ and $\{ t_{k{\bf q}}\}$ are the tunneling matrix elements.

We assume that $T \!\ll\! \Delta$ and, therefore, neglect excited states of the QD concentrating on ground state properties only. We also assume that the QD is weakly coupled to the leads, treating the tunneling Hamiltonian as a perturbation.  Corrections to the position of the peak maximum can be expressed in terms of corrections to the ground state energies of the QD via Eq. (\ref{eq:energy}).  The perturbation series for these corrections contains only even powers as $H_{\rm tun}$ is off-diagonal in the eigenbasis of $H_{0}$.  The $2m^{\rm th}$ correction to the position of the peak is roughly
 \begin{eqnarray}
      2E_C~\overline{{\cal N}^{(2m)}}
   &\approx& \Delta\,\frac{g_L+g_R}{4\pi^2} \,\ln\frac{2E_C}{T}
         \nonumber \\
      &\times&
   \left(\frac{g_L+g_R}{4\pi^2} \,\ln\frac{2E_C}{\Delta}
         \,\ln\frac{2E_C}{T} \right)^{m-1}.
 \end{eqnarray}
Thus, finite-order perturbation theory is applicable if\cite{Kam00}
 \begin{eqnarray}
   \frac{g_L+g_R}{4\pi^2}~ \ln\left(\frac{2E_C}{\Delta}\right)
      \ln\left(\frac{2E_C}{T}\right) \ll 1.
 \end{eqnarray}
To loosen this restriction one should deploy a renormalization group technique which, however, is beyond the scope of this paper.\cite{Kam00,Gla90,And70}

\section{Plan of the Calculation}
\label{sec:approach}

As the exchange interaction constant $J_{\rm s}$ becomes larger, more values of the QD spin $S$ become accessible.  The structure of the corrections to the ground state energies depends on the total QD spin $S$, and this structure becomes very complicated for large values of the spin $S$. Fortunately, for the realistic case $r_{s} = 1.5$, the probability of spin values higher than $\frac{1}{2}$ in an ``odd'' valley is small: $P_2(S \!>\! \frac{1}{2}) \approx 0.01$ for the GUE.  Hence, we can safely assume that in the ``odd'' valley the spin is always equal to $\frac{1}{2}$.  In the ``even'' valley, one has to allow both $S \!=\! 0$ and $S \!=\! 1$ states.

The structure of the expression for the spacing between peaks depends on the allowed spin sequences.  For an ``even'' valley there are only two possible spin sequences:
 \begin{eqnarray}
   {\textstyle \frac{1}{2}} \to 0 \to {\textstyle \frac{1}{2}}~~~~\mbox{and}~~~~
      {\textstyle \frac{1}{2}} \to 1 \to {\textstyle \frac{1}{2}}
         \label{eq:seqeven}
 \end{eqnarray}
where the number in the middle is the spin in the ``even'' valley, while the numbers to the left and right are spin values in the adjacent valleys, Fig.~\ref{fig:su}(b). For an ``odd'' valley there are four possibilities:
 \begin{eqnarray}
   0 \to {\textstyle \frac{1}{2}} \to 0,~~~~ 0 \to {\textstyle \frac{1}{2}} \to 1,
      \nonumber \\
   1 \to {\textstyle \frac{1}{2}} \to 0,~~~~\mbox{and}~~~~ 1 \to {\textstyle \frac{1}{2}} \to 1 \;.
 \end{eqnarray}
To obtain correct expressions for the average spacing between peaks, one should weight these sequences with the appropriate probability of occurrence.

Before proceeding with the calculations, we note several general properties.   First, ensemble-averaged corrections to the ``odd'' and ``even'' spacings are of the same magnitude and opposite sign, Fig.~\ref{fig:su}(b).  Second, the mesoscopic fluctuations of both corrections are equal.  Indeed, the shift in position of an ``even-odd'' ($n \to n \!+\! 1$) peak maximum, Fig.~\ref{fig:su}(b), is determined by the interplay between the $0\!\to\!\frac{1}{2}$ and $1\!\to\!\frac{1}{2}$ spin sequences.  Likewise, the shift of the ``odd-even'' ($n \!-\! 1\to n$) peak is determined by the $\frac{1}{2}\!\to\! 0$ and $\frac{1}{2}\!\to\! 1$ spin sequences.  Now if we sweep the back-gate voltage in the opposite direction and write the same peak as $n\to n \!-\! 1$, then the corresponding spin sequences are exactly the same as they were in the first case: $0\!\to\!\frac{1}{2}$ and $1\!\to\!\frac{1}{2}$.  From this symmetry argument one can conclude that (i) the ensemble-averaged shifts of the ``even-odd'' and ``odd-even'' peaks are of the same magnitude and in the opposite directions\cite{Kam00} and (ii) the mesoscopic fluctuations of both shifts are equal.

Thus, to simplify the calculations we study only the ``even'' spacing case.  This corresponds to the two spin sequences given in Eq.~(\ref{eq:seqeven}).  First, we calculate corrections to the spacing between peaks for both spin sequences.  A complete calculation for the doublet-triplet-doublet spin sequence is in the next section.  Second, we elaborate on how to put these spacings together in the final expression for an ``even'' spacing.  Finally, we calculate GOE and GUE ensemble-averaged corrections to the spacing and the rms fluctuations.

\section{Doublet-Triplet-Doublet Spin Sequence: 
Calculation of the Spacing Between Peaks}
\label{sec:case}

\begin{figure}[b]
\resizebox{.35\textwidth}{!}{\includegraphics{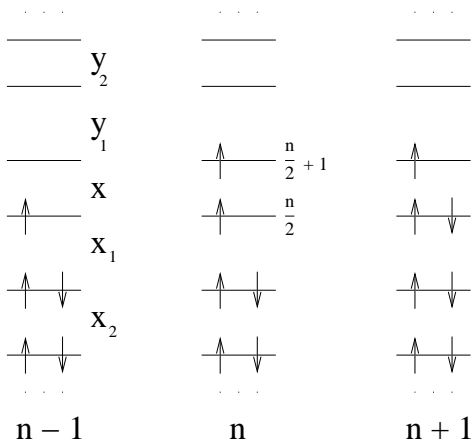}}
\caption{\label{fig:ll}
Occupation of the QD levels in the ground state in three consecutive valleys with total electron number $n-1$, $n$, and $n+1$, respectively.  A doublet-triplet-doublet spin sequence is shown.  The variables $x$, $(x_{1}, x_{2}, \dots)$, and $(y_{1}, y_{2}, \dots)$ denote the energy level spacings in the QD normalized by mean level spacing $\Delta$.  For example, $x \!=\! \left(\varepsilon_{\frac{n}{2}+1}-\varepsilon_{\frac{n}{2}}\right)/\Delta$.} \end{figure}

Let us find the correction to the spacing between peaks for a doublet-triplet-doublet spin sequence.  The corresponding electron occupation of the quantum dot in three consecutive valleys with $n \!-\! 1$, $n$, and $n \!+\! 1$ electrons is shown in Fig.~\ref{fig:ll}.

For the isolated QD, the position of the $n \!-\! 1\to n$ conductance peak maximum is determined by
\begin{eqnarray}
E^{(0)}_{n-1,S=\frac{1}{2}}\big({\cal N}^{(0)}_{n-\frac{1}{2}}\big)=
E^{(0)}_{n,S=1}\big({\cal N}^{(0)}_{n-\frac{1}{2}}\big)
\label{eq:energy0}
\end{eqnarray}
where $E_{n-1,S=\frac{1}{2}}^{(0)}$ and $E_{n,S=1}^{(0)}$ are the ground state energies of the dot Hamiltonian [Eq.~(\ref{eq:dot})].
The corrections due to dot-lead tunneling are different for the doublet and triplet states. The resultant shift in peak position is given by\cite{Kam00}
\begin{equation}
{\cal N}^{(2)}_{n-\frac{1}{2}}=\frac{1}{2E_C}
\left[E^{(2)}_{n,S=1}\big({\cal N}^{(0)}_{n-\frac{1}{2}}\big)
-E^{(2)}_{n-1,S=\frac{1}{2}}\big({\cal N}^{(0)}_{n-\frac{1}{2}}\big)\right] \;.
\label{eq:secondn2}
\end{equation}
Note that for the second-order correction to the position, the ground state energies are taken at the gate voltage obtained in the zeroth-order calculation, Eq.~(\ref{eq:energy0}).

Analogous equations hold for the $n \to n\!+\!1$ conductance peak. The spacing between these two conductance peaks is then defined as 
\begin{equation}
{\cal U}_{n,S=1}=
{\cal N}_{n+\frac{1}{2}}\big(1\to {\textstyle \frac{1}{2}}\big)
-{\cal N}_{n-\frac{1}{2}}\big({\textstyle \frac{1}{2}}\to 1\big).
\label{eq:totalspacing}
\end{equation}

\subsection{Zeroth Order: Isolated Quantum Dot}

For the doublet-triplet $n \!-\! 1\to n$ sequence, Eq.~(\ref{eq:energy0}) gives
\begin{equation}
{\cal N}^{(0)}_{n-\frac{1}{2}}
\big({\textstyle \frac{1}{2}} \to 1\big) = n - \frac{1}{2}
+ \frac{1}{2E_C}
\left( \varepsilon_{\frac{n}{2}+1}-\frac{5}{4}J_{\rm s}
- \frac{T}{2} \ln\frac{3}{2}\right)
\label{eq:zerothn0}
\end{equation}
where the last temperature-dependent term is the entropic correction to the condition of equal energies.\cite{Gla88} For the position of the $n\to n\!+\!1$ peak maximum we obtain
\begin{equation}
{\cal N}^{(0)}_{n+\frac{1}{2}}\big(1 \to {\textstyle \frac{1}{2}}\big)=n+\frac{1}{2}
+\frac{1}{2E_C}\left(
\varepsilon_{\frac{n}{2}}+\frac{5}{4}J_{\rm s}
+\frac{T}{2}\ln\frac{3}{2}\right).
\end{equation}
Thus the spacing between peaks in zeroth order is
\begin{eqnarray}
{\cal U}^{(0)}_{n,S=1}(x)=1+\frac{5j-2x}{2\xi}
+\frac{T}{2E_C}\ln\frac{3}{2},
\label{eq:u0x}
\end{eqnarray}
where $j \!=\! J_{\rm s}/\Delta$ and $x \!=\! \left(\varepsilon_{\frac{n}{2}+1}-\varepsilon_{\frac{n}{2}}\right)/\Delta$  (see Fig.~\ref{fig:ll}).  Similarly, for the doublet-singlet-doublet spin sequence the spacing is
\begin{eqnarray}
{\cal U}^{(0)}_{n,S=0}(x)=1+
\frac{2x-3j}{2\xi}-\frac{T}{2E_C}\ln 2.
\label{eq:u1x}
\end{eqnarray}
Note that in both cases ${\cal U}_{n}^{(0)}$ depends only on the spacing $x$.

\subsection{Second Order: Contribution From Virtual Processes}

\begin{figure}[b]
\resizebox{.42\textwidth}{!}{\includegraphics{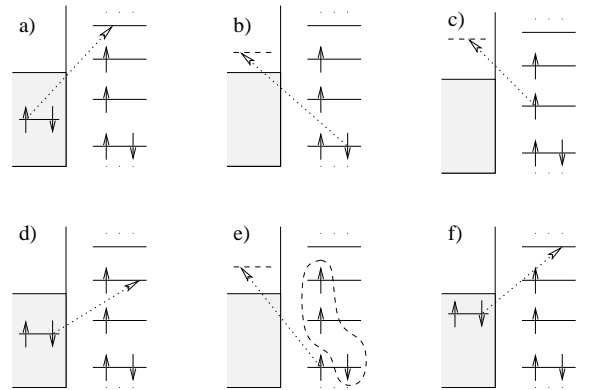}}
\caption{\label{fig:s1} 
Six distinct types of the virtual processes contribute to the ground state energy correction for the QD in the triplet state.  Only tunneling processes in (or out of) the left lead are shown.  In the first four cases spin of the dot in the virtual state ${\cal S}$ has a definite value.  In the last two cases (e) and (f) QD spin in the virtual state has two allowed values: ${\cal S}=\frac{1}{2}$ and ${\cal S}=\frac{3}{2}$ with the probabilities $w_{\cal S}$ given by Eq.~(\ref{eq:ws}).  The electron structure of the virtual state corresponding to two allowed values of ${\cal S}$ is circled by dashed line in panel (e).}
\end{figure}

Let us consider in detail the second-order correction to the ground state energy of the triplet for subsequent use in (\ref{eq:secondn2}):
\begin{equation}
E^{(2)}_{n,S=1}\left({\cal N}\right)={\sum_m}'
\frac{\left|\left<\psi^{(0)}_{m}\left| H_{\rm tun}
\right|\psi^{(0)}_{n,S=1}\right>\right|^{2}}
{E^{(0)}_{n,S=1}\left({\cal N}\right)-E^{(0)}_{m}\left({\cal N}\right)},
\label{eq:e2}
\end{equation}
where the sum is over all possible virtual states.  $E^{(0)}$ and $\left|\psi^{(0)}\right>$ are the eigenvalues and eigenvectors of $H_{\rm dot}$, Eq.~(\ref{eq:dot}).

Different terms in Eq.~(\ref{eq:e2}) have a different structure depending on the type of  virtual state involved; six possibilities are shown in Fig.~\ref{fig:s1}.  To take into account all virtual processes, we sum over all energy levels in the QD and integrate over states in each lead. To simplify the calculation even further, we assume (just for a moment) that $T \!=\! 0$ so that the Fermi distribution in the leads is a step function.  Later, we will see how $T$ reappears as a lower cutoff within a logarithm. 

Following the order of terms in Fig.~\ref{fig:s1}, the second-order correction to the triplet ground state energy is
\begin{widetext}
\begin{eqnarray}
E^{(2)}_{n,S=1}({\cal N})=
-\sum^{\infty}_{k=\frac{n}{2}+2}\sum_{{\bf p}}
\frac{\theta (-\varepsilon_{\bf p})\left|t_{k{\bf p}}\right|^2}
{\left(\varepsilon_k-\varepsilon_{\bf p}\right)
+2E_C(n+\frac{1}{2}-{\cal N})-\frac{7}{4}J_{\rm s}}
&-&\sum_{k=1}^{\frac{n}{2}-1}\sum_{{\bf p}}
\frac{\theta (\varepsilon_{\bf p})\left|t_{k{\bf p}}\right|^2}
{\left(\varepsilon_{\bf p}-\varepsilon_k\right)
-2E_C(n-\frac{1}{2}-{\cal N})-\frac{7}{4}J_{\rm s}}
\nonumber \\
-\sum^{\frac{n}{2}+1}_{k=\frac{n}{2}}\sum_{{\bf p}}
\frac{\theta (\varepsilon_{\bf p})\left|t_{k{\bf p}}\right|^2}
{\left(\varepsilon_{\bf p}-\varepsilon_k\right)
-2E_C(n-\frac{1}{2}-{\cal N})+\frac{5}{4}J_{\rm s}}
&-&\sum_{k=\frac{n}{2}}^{\frac{n}{2}+1}\sum_{{\bf p}}
\frac{\theta (-\varepsilon_{\bf p})\left|t_{k{\bf p}}\right|^2}
{\left(\varepsilon_k-\varepsilon_{\bf p}\right)
+2E_C(n+\frac{1}{2}-{\cal N})+\frac{5}{4}J_{\rm s}}
\nonumber \\
-\sum_{{\cal S}=\frac{1}{2},\frac{3}{2}}w_{\cal S}
\left[
\sum^{\frac{n}{2}-1}_{k=1}\sum_{{\bf p}}
\frac{\theta (\varepsilon_{\bf p})\left|t_{k{\bf p}}\right|^2}
{\left(\varepsilon_{\bf p}-\varepsilon_k\right)
-2E_C(n-\frac{1}{2}-{\cal N})+f_{\cal S}J_{\rm s}}
\right. & + & \left. \sum_{k=\frac{n}{2}+2}^{\infty}\sum_{{\bf p}}
\frac{\theta (-\varepsilon_{\bf p})\left|t_{k{\bf p}}\right|^2}
{\left(\varepsilon_k-\varepsilon_{\bf p}\right)
+2E_C(n+\frac{1}{2}-{\cal N})+f_{\cal S}J_{\rm s}}
\right]
\nonumber \\
&+&\left\{\mbox{similar terms for the right lead:}~{\bf p}\to{\bf q}\right\},
\label{eq:e2t}
\end{eqnarray}
\end{widetext}
where ${\cal S}$ is the spin of the QD in the virtual state.  One can easily find ${\cal S}$ for the first four processes, Figs.~\ref{fig:s1}(a)-(d), and so calculate the denominators for the first four terms in (\ref{eq:e2t}): the values are $\frac{3}{2}$, $\frac{3}{2}$, $\frac{1}{2}$, and $\frac{1}{2}$, respectively.  In the last two cases, Figs.~\ref{fig:s1}(e) and (f), the QD spin in the virtual state can take two values, ${\cal S}=\frac{1}{2}$ or $\frac{3}{2}$; it does so with the following probabilities
\begin{eqnarray}
w_{\frac{1}{2}} = {\textstyle \frac{2}{3}}~~~~\mbox{and}~~~~
w_{\frac{3}{2}} = {\textstyle \frac{1}{3}} \;.
\label{eq:ws}
\end{eqnarray}
The corresponding contributions to the energy correction must be weighted accordingly.  In addition, the energy difference in the denominators depends on ${\cal S}$; to account for this dependence, we introduce an additional function
\begin{eqnarray}
f_{\cal S} \equiv 2-{\cal S}\left({\cal S}+1\right),
\end{eqnarray}
appearing in the denominators of the fifth and sixth terms in (\ref{eq:e2t}).

Let us integrate over the continuous energy levels in the lead. The sum can be replaced by an integral,
$\sum_{\bf p}\,\cdots~\longrightarrow
\int d\varepsilon~\nu_{L}(\varepsilon )\,\cdots\; $,
where $\nu_{L}$ is the density of states in the left lead.  Taking the dot-lead contacts to be point-like, the tunneling matrix elements $\{t_{k{\bf p}}\}$ depend on the momentum in the lead ${\bf p}$ only weakly; hence, $t_{k{\bf p}} \approx t_{kL}$.  In addition, as the leads are formed from 2DEG, their density of states is roughly independent of energy.  We assume that it is constant in the energy band of $2E_C$ near the Fermi surface.  Then the result of integrating over the energy spectrum in the lead (in schematic form) for the first term in Eq.~(\ref{eq:e2t}) is
\begin{eqnarray}
\sum_{\bf p}\frac{\theta(-\varepsilon_{\bf p})
\left|t_{k{\bf p}}\right|^{2}}
{\epsilon_k -\varepsilon_{\bf p}}
~\longrightarrow~
\nu_L\left|t_{kL}\right|^{2}\ln
\left|\frac{\varepsilon}{\epsilon_k}
\right|_{\varepsilon\to\infty}.
\end{eqnarray}
This expression diverges, but when we calculate an {\it observable}, e.g. the shift in the position of the peak maximum [Eq.~(\ref{eq:secondn2})], we encounter the energy difference between corrections to the triplet and doublet states.  The result for the shift is, therefore, finite:
\begin{eqnarray}
\left.\left(
\ln\left|\frac{\varepsilon}{\epsilon_k}\right|
-\ln\left|\frac{\varepsilon}{\epsilon'_k}\right|
\right)\right|_{\varepsilon\to\infty}
=\ln\left|\frac{\epsilon'_k}{\epsilon_k}\right|.
\label{eq:finite}
\end{eqnarray}

In a similar fashion one can calculate the second-order correction to the ground state energy of the doublet.  The difference of these energies at the gate voltage corresponding to the peak maximum in zeroth order [Eq.~(\ref{eq:zerothn0})], needed in  Eq.~(\ref{eq:secondn2}), then follows.  There is one resonant term, proportional to $\ln\left({2E_C}/{T}\right)$,  in which the lower cutoff $T$ appears because of the entropic term in Eq.~(\ref{eq:zerothn0}). 
Alternatively, $T$ would appear as the natural cutoff for the resonant term upon reintroduction of the Fermi-Dirac distribution for the occupation numbers in the leads.

For a point-like dot-lead contact, the tunneling matrix element is proportional to the value of the electron wave function in the QD at the point of contact: $t_{k\alpha}\propto\psi_k({\bf r}_{\alpha})$, where $\alpha =L$ or $R$. Here, we neglect the fluctuations of the electron wave function in the large lead.  Thus, the following identity is valid:
\begin{eqnarray}
\nu_{\alpha}\left|t_{k\alpha}\right|^2
=\frac{\Delta}{4\pi^{2}}~g_{\alpha}
\frac{\left|\psi_{k}\left({\bf r}_{\alpha}\right)\right|^{2}}
{\left<\left|\psi_{k}\left({\bf r}_{\alpha}\right)\right|^{2}\right>},
\label{eq:ttopsi}
\end{eqnarray}
where the average in the denominator is taken over the statistical ensemble.  Note that by taking the ensemble average of both sides of (\ref{eq:ttopsi}), one arrives at the standard golden rule expression for the dimensionless conductance.

In our calculations we take advantage of the fact that
$J_{\rm s} \!<\! \Delta \!\ll\! E_C$ and neglect terms that are of order $1/\xi \!=\! \Delta/2E_C$.
Sums like
\begin{eqnarray}
-\frac{1}{2}\sum_{k=\frac{n}{2}+2}^{\infty}
\ln\left(1+\frac{2J_{\rm s}}
{\varepsilon_{k}-\varepsilon_{\frac{n}{2}+1}}\right)
\end{eqnarray}
are split using
$\sum_{k=\frac{n}{2}+2}^{\infty}\cdots
\!=\! \sum_{k=\frac{n}{2}+2}^{\frac{n}{2}+\xi +1}\cdots
\!+\! \sum_{k=\frac{n}{2}+\xi +2}^{\infty}\cdots$,
and so the last term, which is $O(1/\xi )$, is dropped. Likewise, 
expressions like
\begin{eqnarray}
\frac{2}{3}
\sum_{k=\frac{n}{2}+2}^{\infty}
\ln\left(1+\frac{3}{2}~
\frac{J_{\rm s}}{\varepsilon_{k}-\varepsilon_{\frac{n}{2}+1}+2E_C}
\right)
\end{eqnarray}
are of order $O(1/\xi )$, and so neglected.

Thus, for the second-order correction to the position of the peak maximum, we obtain
\begin{widetext}
\begin{eqnarray}
{\cal N}^{(2)}_{n-\frac{1}{2}}\big({\textstyle \frac{1}{2}} \to 1\big)
=\frac{1}{4\pi^{2}\xi}\sum_{\alpha =L,R}
g_{\alpha}&&
\left[
-
2\sum_{k\ne\frac{n}{2},\frac{n}{2}+1}
\mbox{sign}\left(\frac{n}{2}-k\right)
\frac{\left|\psi_{k}\left({\bf r}_{\alpha}\right)\right|^{2}}
{\left<\left|\psi_{k}\left({\bf r}_{\alpha}\right)\right|^{2}\right>}
\ln\left(\frac{2E_C}{\left|\varepsilon_{\frac{n}{2}+1}-\varepsilon_{k}\right|}+1\right)
\right.
\nonumber \\
\left.
-\frac{\left|\psi_{\frac{n}{2}}\left({\bf r}_{\alpha}\right)\right|^{2}}
{\left<\left|\psi_{\frac{n}{2}}\left({\bf r}_{\alpha}\right)\right|^{2}\right>}
\ln\left(\frac{2J_{\rm s}}{\varepsilon_{\frac{n}{2}+1}-\varepsilon_{\frac{n}{2}}}-1\right)
\right.
&+&
\left.
\frac{1}{2}
\frac{\left|\psi_{\frac{n}{2}+1}\left({\bf r}_{\alpha}\right)\right|^{2}}
{\left<\left|\psi_{\frac{n}{2}+1}\left({\bf r}_{\alpha}\right)\right|^{2}\right>}
\left(\ln\frac{E_C}{J_{\rm s}}+\ln\frac{2E_C}{T}\right)
\right.
\nonumber \\
\left.
+\frac{4}{3}
\sum_{k=\frac{n}{2}-\xi}^{\frac{n}{2}-1}
\frac{\left|\psi_{k}\left({\bf r}_{\alpha}\right)\right|^{2}}
{\left<\left|\psi_{k}\left({\bf r}_{\alpha}\right)\right|^{2}\right>}
\ln\left|1-\frac{3J_{\rm s}}{\varepsilon_{\frac{n}{2}+1}-\varepsilon_{k}}\right|
\right.
&-&
\left.
\frac{1}{2}
\sum_{k=\frac{n}{2}+2}^{\frac{n}{2}+\xi}
\frac{\left|\psi_{k}\left({\bf r}_{\alpha}\right)\right|^{2}}
{\left<\left|\psi_{k}\left({\bf r}_{\alpha}\right)\right|^{2}\right>}
\ln\left(1+\frac{2J_{\rm s}}{\varepsilon_{k}-\varepsilon_{\frac{n}{2}+1}}\right)
+O\left(\frac{1}{\xi}\right)
\right],
\end{eqnarray}
where $2J_{\rm s}>\varepsilon_{\frac{n}{2}+1} \!-\! \varepsilon_{\frac{n}{2}}>0$ because the total spin of the QD with $n$ electrons is equal to 1.

In a similar fashion one can find the shift in the position of the other peak maximum, ${\cal N}^{(2)}_{\frac{n}{2}+1}\left(1\to\frac{1}{2}\right)$.  Then, according to (\ref{eq:totalspacing}), the difference of these two shifts yields the second-order correction to the spacing for the doublet-triplet-doublet spin sequence:
\begin{eqnarray}
{\cal U}_{n,S=1}^{(2)}
&=&\frac{1}{4\pi^{2}\xi}\sum_{\alpha =L,R}
g_{\alpha}
\left\{
2\sum_{k\ne\frac{n}{2},\frac{n}{2}+1}
\mbox{sign}\left(\frac{n}{2}-k\right)
\frac{\left|\psi_{k}\left({\bf r}_{\alpha}\right)\right|^{2}}
{\left<\left|\psi_{k}\left({\bf r}_{\alpha}\right)\right|^{2}\right>}
\left[
\ln\left(\frac{2E_C}{\left|\varepsilon_{\frac{n}{2}+1}-\varepsilon_{k}\right|}+1\right)
-\ln\left(\frac{2E_C}{\left|\varepsilon_{\frac{n}{2}}-\varepsilon_{k}\right|}+1\right)
\right]
\right.
\nonumber \\
&+&
\left.
\left(
\frac{\left|\psi_{\frac{n}{2}}\left({\bf r}_{\alpha}\right)\right|^{2}}
{\left<\left|\psi_{\frac{n}{2}}\left({\bf r}_{\alpha}\right)\right|^{2}\right>}
+\frac{\left|\psi_{\frac{n}{2}+1}\left({\bf r}_{\alpha}\right)\right|^{2}}
{\left<\left|\psi_{\frac{n}{2}+1}\left({\bf r}_{\alpha}\right)\right|^{2}\right>}
\right)
\left[
\ln\left(\frac{2J_{\rm s}}{\varepsilon_{\frac{n}{2}+1}-\varepsilon_{\frac{n}{2}}}-1\right)
-\frac{1}{2}\ln\frac{E_C}{J_{\rm s}}
-\frac{1}{2}\ln\frac{2E_C}{T}
\right]
\right.
\nonumber \\
&+&
\left.
\sum_{k=\frac{n}{2}-\xi}^{\frac{n}{2}-1}
\frac{\left|\psi_{k}\left({\bf r}_{\alpha}\right)\right|^{2}}
{\left<\left|\psi_{k}\left({\bf r}_{\alpha}\right)\right|^{2}\right>}
\left[
\frac{1}{2}
\ln\left(1+\frac{2J_{\rm s}}{\varepsilon_{\frac{n}{2}}-\varepsilon_{k}}\right)
-\frac{4}{3}
\ln\left|1-\frac{3J_{\rm s}}{\varepsilon_{\frac{n}{2}+1}-\varepsilon_{k}}\right|
\right]
\right.
\nonumber \\
&+&
\left.
\sum_{k=\frac{n}{2}+2}^{\frac{n}{2}+\xi}
\frac{\left|\psi_{k}\left({\bf r}_{\alpha}\right)\right|^{2}}
{\left<\left|\psi_{k}\left({\bf r}_{\alpha}\right)\right|^{2}\right>}
\left[
\frac{1}{2}
\ln\left(1+\frac{2J_{\rm s}}{\varepsilon_{k}-\varepsilon_{\frac{n}{2}+1}}\right)
-\frac{4}{3}
\ln\left|1-\frac{3J_{\rm s}}{\varepsilon_{k}-\varepsilon_{\frac{n}{2}}}\right|
\right]
+O\left(\frac{1}{\xi}\right)
\right\}.
\label{eq:u2triplet}
\end{eqnarray}
A potential complication is that the addition of two electrons to the quantum dot ($n \!-\! 1\to n\to n \!+\! 1$) may scramble the energy levels and wave functions of the QD.\cite{Bla97,Val98,Usa02} Since the number of added electrons is small, we assume that the same realization of the Hamiltonian, Eq.~(\ref{eq:hamilt}), is valid in all three valleys.\cite{Pat98}

For the second-order correction to the spacing for the doublet-singlet-doublet spin sequence we similarly obtain
\begin{eqnarray}
{\cal U}_{n,S=0}^{(2)}
&=&\frac{1}{4\pi^{2}\xi}\sum_{\alpha =L,R}
g_{\alpha}
\left\{
-2\sum_{k\ne\frac{n}{2},\frac{n}{2}+1}
\mbox{sign}\left(\frac{n}{2}-k\right)
\frac{\left|\psi_{k}\left({\bf r}_{\alpha}\right)\right|^{2}}
{\left<\left|\psi_{k}\left({\bf r}_{\alpha}\right)\right|^{2}\right>}
\left[
\ln\left(\frac{2E_C}{\left|\varepsilon_{\frac{n}{2}+1}-\varepsilon_{k}\right|}+1\right)
-\ln\left(\frac{2E_C}{\left|\varepsilon_{\frac{n}{2}}-\varepsilon_{k}\right|}+1\right)
\right]
\right.
\nonumber \\
&+&
\left.
\left(
\frac{\left|\psi_{\frac{n}{2}}\left({\bf r}_{\alpha}\right)\right|^{2}}
{\left<\left|\psi_{\frac{n}{2}}\left({\bf r}_{\alpha}\right)\right|^{2}\right>}
+\frac{\left|\psi_{\frac{n}{2}+1}\left({\bf r}_{\alpha}\right)\right|^{2}}
{\left<\left|\psi_{\frac{n}{2}+1}\left({\bf r}_{\alpha}\right)\right|^{2}\right>}
\right)
\left[
\frac{3}{2}
\ln\left(1-\frac{2J_{\rm s}}{\varepsilon_{\frac{n}{2}+1}-\varepsilon_{\frac{n}{2}}}\right)
-2\ln\left(
\frac{2E_C}{\varepsilon_{\frac{n}{2}+1}-\varepsilon_{\frac{n}{2}}}+1
\right)
+\ln\frac{2E_C}{T}
\right]
\right.
\nonumber \\
&+&
\left.
\frac{3}{2}
\sum_{k=\frac{n}{2}-\xi}^{\frac{n}{2}-1}
\frac{\left|\psi_{k}\left({\bf r}_{\alpha}\right)\right|^{2}}
{\left<\left|\psi_{k}\left({\bf r}_{\alpha}\right)\right|^{2}\right>}
\ln\left(1-\frac{2J_{\rm s}}{\varepsilon_{\frac{n}{2}+1}-\varepsilon_{k}}\right)
+\frac{3}{2}
\sum_{k=\frac{n}{2}+2}^{\frac{n}{2}+\xi}
\frac{\left|\psi_{k}\left({\bf r}_{\alpha}\right)\right|^{2}}
{\left<\left|\psi_{k}\left({\bf r}_{\alpha}\right)\right|^{2}\right>}
\ln\left(1-\frac{2J_{\rm s}}{\varepsilon_{k}-\varepsilon_{\frac{n}{2}}}\right)
+O\left(\frac{1}{\xi}\right)
\right\},
\label{eq:u2singlet}
\end{eqnarray}
\end{widetext}
where $\varepsilon_{\frac{n}{2}+1} \!-\! \varepsilon_{\frac{n}{2}} > 2J_{\rm s} \ge 0$ because the total spin of the QD with $n$ electrons is equal to $0$ in this case.

Unlike the zeroth-order spacings, the second-order corrections are functions of many energy level spacings as well as the wave functions at the dot-lead contact points:
${\cal U}_{n,S}^{(2)}=
{\cal U}_{n,S}^{(2)}(x,{\bf X},{\bf Y};
\{Z_{k\alpha}\})$,
where $x$, ${\bf X} = (x_{1}, x_{2}, \dots )$, and ${\bf Y} = (y_{1}, y_{2}, \dots )$ are the energy level spacings in the QD normalized by the mean level spacing $\Delta$ (see Fig.~\ref{fig:ll}) and 
\begin{eqnarray}
Z_{k\alpha} \equiv
\frac{\left|\psi_{k}\left({\bf r}_{\alpha}\right)\right|^{2}}
{\left<\left|\psi_{k}\left({\bf r}_{\alpha}\right)\right|^{2}\right>} \;.
\end{eqnarray}

The expressions for ${\cal U}^{(2)}$ suggest that the main contribution to their fluctuation comes from the fluctuation of the energy level $x$ and the wave functions $\left\{\psi_{k}\left({\bf r}_{\alpha} \right) \right\}$.  The other spacings, ${\bf X}$ and ${\bf Y}$, always appear within a logarithm; therefore, their contribution to the fluctuation of ${\cal U}^{(2)}$ is small.  With good accuracy, one can replace these levels by their mean value 
\begin{equation}
{\cal U}_{n,S}^{(2)} \approx
{\cal U}_{n,S}^{(2)}(x,{\bf 1},{\bf 1};
\{Z_{k\alpha}\}) \equiv {\cal U}_{n,S}^{(2)}(x;\{Z_{k\alpha}\}) \;.
\end{equation}
\newpage\noindent 
Converting to dimensionless units, we find that
\begin{equation}
{\cal U}^{(2)}_{n,S}\left(x;\left\{Z_{k\alpha}\right\}\right)
=\frac{1}{4\pi^{2}\xi}
\sum_{\alpha =L,R}g_{\alpha}\Phi_{\alpha ,S}
\left(x;\left\{Z_{k\alpha}\right\}\right),
\label{eq:u2gen}
\end{equation}
where
\begin{widetext}
\begin{eqnarray}
\Phi_{\alpha ,S=0}
\left(x;\left\{Z_{k\alpha}\right\}\right)
=\left(Z_{\frac{n}{2},\alpha}+Z_{\frac{n}{2}+1,\alpha}\right)
\left[
-\ln \xi +\ln\delta +\frac{1}{2}\ln x+\frac{3}{2}\ln (x-2j)
\right]
+2\sum_{l=1}^{\infty}
\left(Z_{\frac{n}{2}-l,\alpha}+Z_{\frac{n}{2}+1+l,\alpha}\right)
\nonumber \\
\times
\left[
\ln\left(1+\frac{x}{l}\right)-\ln\left(1+\frac{x}{\xi +l}\right)
\right]
+\frac{3}{2}\sum_{l=1}^{\xi}
\left(Z_{\frac{n}{2}-l,\alpha}+Z_{\frac{n}{2}+1+l,\alpha}\right)
\ln\left(1-\frac{2j}{x+l}\right)
+O\left(1\right),
\label{eq:ups0}
\end{eqnarray}
\begin{eqnarray}
\Phi_{\alpha ,S=1}
\left(x;\left\{Z_{k\alpha}\right\}\right)
=\left(Z_{\frac{n}{2},\alpha}+Z_{\frac{n}{2}+1,\alpha}\right)
\left[
-\ln \xi -\frac{1}{2}\ln\delta +\frac{1}{2}\ln 2j+\ln\left(\frac{2j}{x}-1\right)
\right]
-2\sum_{l=1}^{\infty}
\left(Z_{\frac{n}{2}-l,\alpha}+Z_{\frac{n}{2}+1+l,\alpha}\right)
\nonumber \\
\times
\left[
\ln\left(1+\frac{x}{l}\right)-\ln\left(1+\frac{x}{\xi +l}\right)
\right]
+\sum_{l=1}^{\xi}
\left(Z_{\frac{n}{2}-l,\alpha}+Z_{\frac{n}{2}+1+l,\alpha}\right)
\left[
\frac{1}{2}\ln\left(1+\frac{2j}{l}\right)
-\frac{4}{3}\ln\left|1-\frac{3j}{x+l}\right|
\right]
+O\left(1\right),
\label{eq:ups1}
\end{eqnarray}
\end{widetext}
where $\delta =\Delta /T$.  Here, the upper limit in two of the sums is infinity because the Fermi energy is the largest energy scale.

In summary, the total spacing is
\begin{eqnarray}
{\cal U}_{n,S}={\cal U}_{n,S}^{(0)}(x)+
{\cal U}_{n,S}^{(2)}(x;\{Z_{k\alpha}\}),
\end{eqnarray}
where the first term is given by Eqs.~(\ref{eq:u0x})-(\ref{eq:u1x}) and the second by (\ref{eq:u2gen})-(\ref{eq:ups1}).  The spin of the QD in the even valley, $S$, can take two values, $0$ or $1$, depending on the spacing $x$.

\section{Ensemble-Averaged Correction to the Peak Spacing}
\label{sec:averaged}

The average and rms correction to the peak spacing can now be found by using the known distribution of the single-particle quantities $x$ and $\{Z_{k\alpha}\}$.
In what follows, $\left<{\cal U}\right>$ denotes the average over the wave functions, $\overline{\cal U}$ denotes the full average over both wave functions and energy levels, and $P(x)$ is the distribution of the spacing $x$.
Since $\left<Z_{k\alpha}\right> \!=\! 1$,
$\left<\Phi_{\alpha,S}\right>$ does not depend on $\alpha$, and the average ``even'' spacing is\cite{Border}
\begin{eqnarray}
\overline{{\cal U}_n^{(2)}} &=& \int_{0}^{\infty}dx\, P(x)\left<{\cal U}_{n,S}^{(2)}\right>
\label{eq:aveugen}
\\
& = &
\frac{g_L+g_R}{4\pi^2\xi}
\left(
\int_{2j}^{\infty}dx~P(x)\left<\Phi_{\alpha,S=0}^{(2)}\right>
\right.
\\
& & 
\left. \qquad
+\int_{0}^{2j}dx~P(x)\left<\Phi_{\alpha,S=1}^{(2)}\right>
\right) \;.
\nonumber
\end{eqnarray}
Using the asymptotic formulas
\begin{eqnarray}
\sum\limits_{l=1}^{\infty} & &  \!\!\Big[
\ln\Big(1+\frac{x}{l}\Big)
-\ln\Big(1+\frac{x}{\xi +l}\Big)
\Big]\approx x\,\ln \xi ,
\nonumber \\
\sum\limits_{l=1}^{\xi} & &
\ln\Big(1-\frac{2j}{x+l}\Big)\approx - 2 j\,\ln \xi 
\end{eqnarray}
for $\xi \!\gg\! 1$ in the expressions for $\left<\Phi_{\alpha,S}\right>$, we find
\begin{eqnarray}
\left<\Phi_{\alpha ,S=0}\right> 
& = & 2\, (2x-3j-1)\,\ln \xi +2\,\ln\delta
\nonumber
\\
\left<\Phi_{\alpha ,S=1}\right> 
& = & 2\, (-2x+5j-1)\,\ln \xi -\ln\delta \;,
\label{eq:aveups}
\end{eqnarray}
valid for $\xi ,\delta\gg 1$.  By carrying out the integration over the distribution of the spacing $x$, the final expression is
\begin{eqnarray}
\overline{{\cal U}^{(2)}_{n}\left(j\right)}
&=& \frac{g_{L}+g_{R}}{4\pi^{2}\xi}
\big[{\cal C}(j)\ln \xi +{\cal D}(j)\ln\delta +O(1)\big]
\nonumber \\
{\cal C}(j)&=&2\big[-8jP_{0}(2j)+4x_{0}(2j)+5j-3\big],
\label{eq:even2} \\
{\cal D}(j)&=&3P_{0}(2j)-1,
\nonumber
\end{eqnarray}
where $P_{0}(2j) = \int_{2j}^{\infty}dx\, P(x)$ and $x_{0}(2j) = \int_{2j}^{\infty}dx\, x\, P(x)$.  Note that $P_{0}(2j)$ is the probability of  obtaining a singlet ground state while $x_0(2j)/P_0(2j)$ is the average value of $x$ given that the ground state is a singlet.

For the CI model, $j \!=\!0$ and, hence, ${\cal C}(0)\!=\!{\cal D}(0)\!=\!2$. 
In this limit, then, the ensemble-averaged correction to the spacing is
\begin{eqnarray}
\overline{{\cal U}^{(2)}_{n}\left(0\right)}
=\frac{g_L+g_R}{2\pi^{2}\xi}\ln\frac{2E_C}{T} \;,
\label{eq:un20}
\end{eqnarray}
in agreement with previous work.\cite{Kam00} The magnitude here is approximately $0.05 \left(g_L+g_R\right)\ln\left(2E_C/T\right)$
in units of the mean level spacing.

It is convenient to relate the average change in spacing at non-zero $J_{\rm s}$ to that at $J_{\rm s} \!=\! 0$:
\begin{eqnarray}
\delta u(j)
& \equiv & \frac{\;\overline{{\cal U}^{(2)}_{n}\left(j\right)}\;}
{\overline{{\cal U}^{(2)}_{n}\left(0\right)}}
= \frac{\lambda\,{\cal C}(j) + {\cal D}(j)}{2(\lambda+1)},
\label{eq:ave2plot}
\\
\lambda & \equiv & \frac{\ln \xi}{\ln\delta}
=\frac{\ln(2E_C / \Delta)}{\ln (\Delta /T)} \;.
\label{eq:lambda}
\end{eqnarray}
The dependence of $\delta u$ on $j$ is fully determined by the parameter $\lambda$ and the choice of the random matrix ensemble.

\begin{figure}[t]
\resizebox{.4\textwidth}{!}{\includegraphics{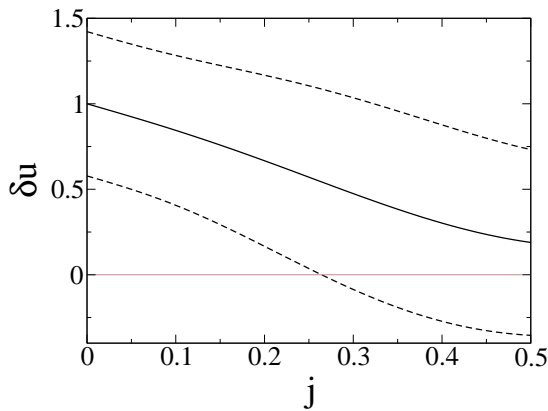}}
\caption{\label{fig:ave1}
Correction to ``even'' peak spacing as a function of strength of exchange, 
$j \!=\! J_{\rm s}/\Delta$, normalized by the correction at $j \!=\! 0$. GUE case with $\lambda \!=\! 1$ and $g_{L} \!=\! g_{R}$.
Solid: Ensemble average. Dashed: Ensemble average
plus/minus the rms, showing the width of the distribution.
}
\end{figure}

\begin{figure}[b]
\resizebox{.4\textwidth}{!}{\includegraphics{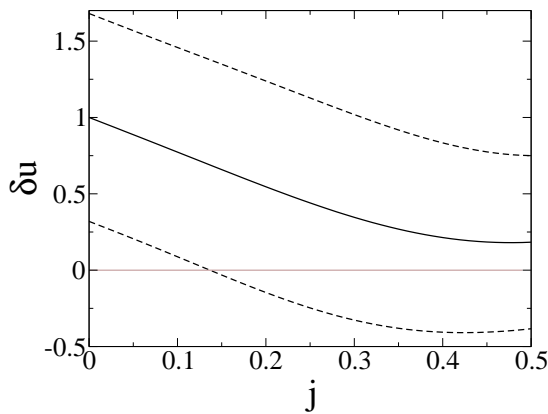}}
\caption{\label{fig:ave3}
The same quantities as in Fig.~\ref{fig:ave1} 
plotted for $\lambda = 3$.}
\end{figure}

\begin{figure}[t]
\resizebox{.4\textwidth}{!}{\includegraphics{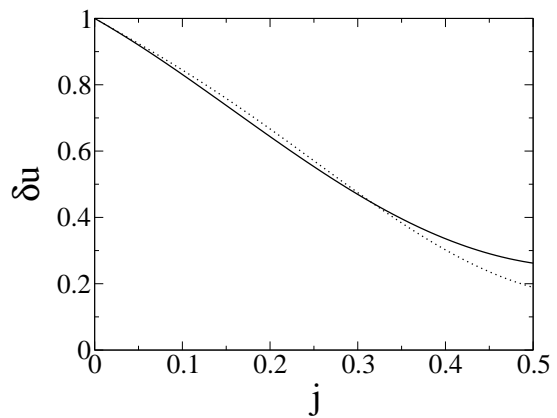}}
\caption{\label{fig:goe}
Ensemble-averaged correction to the ``even'' peak spacing as a function of
strength of exchange, $j \!=\! J_{\rm s}/\Delta$, normalized by the correction at $j \!=\! 0$ for $\lambda \!=\! 1$. Solid: GOE.  Dotted: GUE.}
\end{figure}

Figures~\ref{fig:ave1} and \ref{fig:ave3} show the results in the GUE ensemble for $\lambda \!=\! 1$ and $3$, respectively, and Fig.~\ref{fig:goe} shows those for the GOE ensemble at $\lambda \!=\! 1$.
In evaluating these expressions, we use the Wigner surmise distributions for $P(x)$, which allow an analytic evaluation of $P_{0}(2j)$ and $x_{0}(2j)$.
As $j$ increases, the average correction to the peak spacing decreases monotonically in all three cases.  (Note, however, that our results are not completely trustworthy when $0.4 \!<\! j \!<\! 0.5$ because in this regime higher spin states should be taken into account.)  Since $\lambda$ depends on $\xi$ and $\delta$ only logarithmically, the qualitative behavior of $\delta u (j)$ is very robust with respect to changes in charging energy, mean level spacing, or temperature.

\begin{figure}[b]
\resizebox{.4\textwidth}{!}{\includegraphics{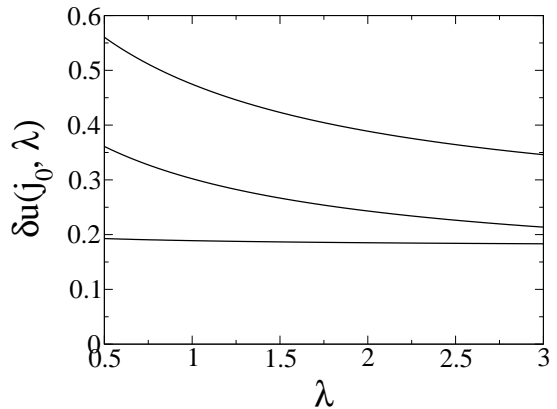}}
\caption{\label{fig:fj} 
GUE ensemble-averaged correction to the even peak spacing as a function of $\lambda \!=\! \ln(2E_C / \Delta) / \ln (\Delta /T)$ at $j \!=\! j_{0}$.  The curves correspond to $j_{0} = 0.3$, $0.4$, and $0.5$ from the top to the bottom.}
\end{figure}

Similarly, the dependence of $\delta u$ on $\lambda$ at $j \!=\! j_{0}$ is
\begin{eqnarray}
\delta u(j_{0},\lambda )
= \frac{\lambda\, {\cal C}(j_{0})+{\cal D}(j_{0})}{2(\lambda+1)} \;.
\label{eq:avej0}
\end{eqnarray}
Figure~\ref{fig:fj} shows results in the GUE case for several values of $j_{0}$.  \textit{Thus, for the realistic value $j_{0} \!=\! 0.3$, the CEI model gives an average correction to the peak spacing that is two to three times smaller than the CI model.}

\section{RMS of The Correction to Peak Spacing due to Mesoscopic Fluctuations}
\label{sec:fluctuations}

Mesoscopic fluctuations of the correction to the peak spacing are characterized by the variance of ${\cal U}^{(2)}$. 
It is convenient to separate the average over the wave functions from that over the spacing $x$, writing
\begin{eqnarray}
\mbox{var}\big({\cal U}_{n}^{(2)}\big)
= \sigma^{2}_{Z}\big({\cal U}_{n}^{(2)}\big)
+ \sigma^{2}_{x}\big({\cal U}_{n}^{(2)}\big),
\label{eq:sigmamain}
\end{eqnarray}
where
\begin{equation}
\sigma^{2}_{Z}
= \int_{0}^{\infty}dx\, P(x)
\Big<\Big(
{\cal U}_{n,S}^{(2)}-\big<{\cal U}_{n,S}^{(2)}\big>
\Big)^{2}\Big>
\label{eq:sigmapsi}
\end{equation}
is the contribution due to wave function fluctuations and 
\begin{equation}
\sigma^{2}_{x}
= \int_{0}^{\infty}\!\!dx\, P(x)
\big<{\cal U}_{n,S}^{(2)}\big>^{2}
-\left[
\int_{0}^{\infty}\!\!dx\, P(x)
\big<{\cal U}_{n,S}^{(2)}\big>
\right]^{2}
\label{eq:sigmax}
\end{equation}
is the contribution due to fluctuation of the spacing $x$.

We start by considering the fluctuations of the wave functions.  As the number of electrons in the dot is large, the distance between the left and right dot-lead contacts is large, $\left|{\bf r}_{L}-{\bf r}_{R}\right|\gg\lambda_F$, where $\lambda_{F}$ is the Fermi wavelength.  Therefore, the wave functions at ${\bf r}_{L}$ and ${\bf r}_{R}$ are uncorrelated,\cite{Pri95}
\begin{eqnarray}
\left<\left( Z_{kL} - 1 \right)
\left( Z_{k'R} - 1 \right)\right> = 0
\end{eqnarray}
for all $k$ and $k'$. The fluctuation of ${\cal U}^{(2)}$ can then be written entirely in terms of the properties of a single lead:
\begin{equation}
\Big<\Big(
{\cal U}_{n,S}^{(2)}-\big<{\cal U}_{n,S}^{(2)}\big>
\Big)^{2}\Big>
=\frac{g_L^2+g_R^2}{\left(4\pi^{2}\xi\right)^2}
\Big<\big(
\Phi_{L,S}-\left<\Phi_{L,S}\right>
\big)^{2}\Big> \;.
\label{eq:sigups}
\end{equation}
The cross terms disappear here because, according to RMT, wave functions of different energy levels are uncorrelated even at the same point in space,
\begin{eqnarray}
\big<
\left( Z_{kL} - 1 \right)
\left( Z_{k'L} - 1 \right)
\big>
=\frac{2}{\beta}\,\delta_{kk'}
\label{eq:zz}
\end{eqnarray}
where $\beta \!=\! 1$ ($\beta \!=\! 2$) for the GOE (GUE) case. In fact, only the $k \!=\! \frac{n}{2}$ and $k \!=\! \frac{n}{2} \!+\! 1$ terms contribute, as one can see by using
\begin{eqnarray}
\sum_{l=1}^{\xi}\,\ln^{2}
\Big( 1 - \frac{2j}{x+l}\Big) = O(1)
\end{eqnarray}
valid for $\xi\gg 1$. Integrating (\ref{eq:sigups}) over the distribution of $x$ according to Eq.~(\ref{eq:sigmapsi}) (keeping in mind $\xi ,\delta \!\gg\! 1$), we obtain
\newpage\noindent
\begin{eqnarray}
\label{eq:sigmazc}
\lefteqn{ 
\sigma_Z^2\Big({\cal U}_n^{(2)}\Big)
=\frac{g_L^2+g_R^2}{\beta\left(4\pi^{2}\xi\right)^2}
\;\times 
}
\\ &
\big\{
4\ln^{2}\!\xi
+\left[3P_{0}(2j)+1\right]\ln^{2}\!\delta
-4\left[3P_0(2j)-1\right]\ln\xi\ln\delta
\big\}.
\nonumber
\end{eqnarray}

In the contribution to the variance due to fluctuation of the level spacing $x$, Eq.~(\ref{eq:sigmax}), $\big<{\cal U}_{n,S}^{(2)}\big>$ can be taken from the previous section. Since the average eliminates the dependence on the lead $\alpha$, we have immediately
\begin{eqnarray}
\big<{\cal U}^{(2)}_{n,S}\big>^{2}
= \left(\frac{g_{L}+g_{R}}{4\pi^{2}\xi}\right)^{2}
\left<\Phi_{L,S}\right>^{2}
\label{eq:uns22}
\end{eqnarray}
where $\left<\Phi_{L,S=0}\right>$ and $\left<\Phi_{L,S=1}\right>$ for $\xi ,\delta\gg 1$ are given by Eq.~(\ref{eq:aveups}).  Using these expressions in Eq.~(\ref{eq:sigmax}), we obtain
\begin{equation}
\sigma_x^2
=\left(\frac{g_{L}+g_{R}}{4\pi^{2}\xi}\right)^{2}
\left(
{\cal C}_{\xi\xi}\ln^{2}\xi
+{\cal C}_{\delta\delta}\ln^{2}\delta
+{\cal C}_{\xi\delta}\ln \xi\ln\delta
\right).
\label{eq:sigmaxnoc}
\end{equation}
Explicit expressions for the coefficients are given below once we reach the final result.

The dependence on $g_{L}$ and $g_{R}$  of the two contributions to the variance is different. In particular, the contribution due to fluctuations of the wave functions [Eq.~(\ref{eq:sigmazc})] is proportional to
\begin{equation}
g_{L}^{2} + g_{R}^{2}
= \frac{\left( g_{L} + g_{R}\right)^{2}}{2}
+ \frac{\left( g_{L} - g_{R}\right)^{2}}{2} \;.
\end{equation}
The first term has the same form as the contribution (\ref{eq:sigmaxnoc}) from fluctuations of $x$. It is convenient to write the total variance as a sum of symmetric and antisymmetric parts. Our final result for the variance is
\begin{eqnarray}
\mbox{var}\big({\cal U}_{n}^{(2)}\big)
= \sigma_{s}^{2}\big({\cal U}_{n}^{(2)}\big)
+ \sigma_{a}^{2}\big({\cal U}_{n}^{(2)}\big)
\label{eq:varianceref}
\end{eqnarray}
where 
\begin{widetext}
\begin{eqnarray}
\sigma_{s}^{2}\left({\cal U}_{n}^{(2)}\right)
& = & \left(\frac{g_L+g_R}{4\pi^{2}\xi}\right)^{2}
\Big(
{\cal S}_{\xi\xi}\ln^{2}\xi
+{\cal S}_{\delta\delta}\ln^{2}\delta
+{\cal S}_{\xi\delta}\ln \xi\ln\delta
\Big)
\label{eq:sigmasym7}
\\
\sigma_{a}^{2}\left({\cal U}_{n}^{(2)}\right)
& =& \left(\frac{g_L-g_R}{4\pi^{2}\xi}\right)^{2}
\Big(
{\cal A}_{\xi\xi}\ln^{2}\xi
+{\cal A}_{\delta\delta}\ln^{2}\delta
+{\cal A}_{\xi\delta}\ln \xi\ln\delta
\Big)
\label{eq:sigmaasym5}
\end{eqnarray}
with the coefficients 
$\{ {\cal S} \}$ and $\{ {\cal A} \}$
given by
\begin{eqnarray}
{\cal S}_{\xi\xi}(j)
&=&\frac{2}{\beta}+16\left({\chi}-1\right)
+64\left[2jP_{0}(2j)-x_{0}(2j)\right]
\big\{ 2j\left[1-P_{0}(2j)\right]-\left[1-x_{0}(2j)\right]\big\},
\label{eq:s1j}
\\
{\cal S}_{\delta\delta}(j)
&=&\frac{1}{2\beta}\left[3P_{0}(2j)+1\right]
+9P_{0}(2j)\left[1-P_{0}(2j)\right],
\\
{\cal S}_{\xi\delta}(j)
&=&-\frac{2}{\beta}\left[3P_{0}(2j)-1\right]
+24\big\{
x_{0}(2j)\left[1-P_{0}(2j)\right]
+\left[1-x_{0}(2j)\right]P_{0}(2j)
-4jP_{0}(2j)\left[1-P_{0}(2j)\right]
\big\},
\\
{\cal A}_{\xi\xi}(j)
&=&\frac{2}{\beta},
~~~~{\cal A}_{\delta\delta}(j)
=\frac{1}{2\beta}\left[3P_{0}(2j)+1\right],
~~~~\mbox{and}~~~~{\cal A}_{\xi\delta}(j)
= - \frac{2}{\beta}\left[3P_{0}(2j)-1\right].
\end{eqnarray}
\newpage\noindent
\end{widetext}
The constant $\chi$ 
introduced in Eq.~(\ref{eq:s1j}) is
\begin{eqnarray}
{\chi} = \int_{0}^{\infty}\!\!dx\,x^{2}\,P(x).
\label{eq:xi}
\end{eqnarray}

For the CI model, $j=0$
and $P_{0}(0)=x_{0}(0)=1$; hence
\begin{eqnarray}
\left.\mbox{var}\big({\cal U}_{n}^{(2)}\big)\right|_{j=0}
=\frac{4}{\beta}\frac{g_L^2+g_R^2}{\left(4\pi^{2}\xi\right)^{2}}
\left(\ln \xi -\ln\delta\right)^{2}
\nonumber \\
+16\left({\chi}-1\right)
\left(\frac{g_L+g_R}{4\pi^{2}\xi}\right)^{2}\ln^{2}\xi \;.
\end{eqnarray}
The first term is due to fluctuation of the wave functions at the dot-lead contacts, and the second term comes from the fluctuation of the level spacing $x$.  The presence of the second term was missed in previous work [see Eq.~(44b) in Ref.~\onlinecite{Kam00}]. If $\xi \!=\! \delta$, the first term vanishes; nonetheless, due to the second term the variance is always finite.  

\begin{figure}[b]
\resizebox{.4\textwidth}{!}{\includegraphics{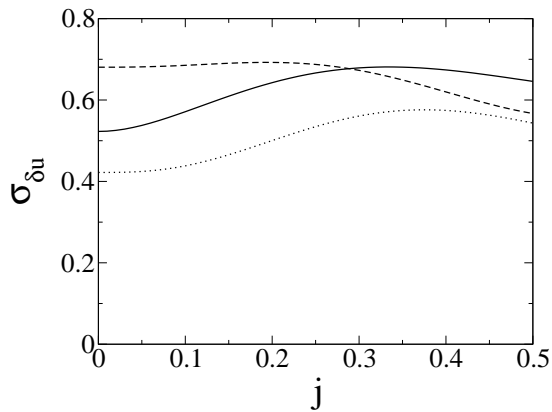}}
\caption{\label{fig:sigma}
The rms of the correction to the peak spacing
for the symmetric setup $g_{L} = g_{R}$
as a function of $j = J_{\rm s}/\Delta$
normalized by the ensemble-averaged correction at $j = 0$,
Eq.~(\ref{eq:sigmasglgr}).
The solid (dotted) curve corresponds to the GOE (GUE)
at $\lambda = 1$.
The dashed curve corresponds to the GUE
at $\lambda=3$.}
\end{figure}

Let us consider a realistic special case of symmetric tunnel barriers, $g_{L}=g_{R}$.\cite{Mau99} Then the asymmetric contribution vanishes, and the rms fluctuation of the correction to the peak spacing normalized by the average correction at $j = 0$ [Eq.~(\ref{eq:un20})] is
\begin{equation}
\sigma_{\delta u}(j)
= \frac{\sigma_{s}\big({\cal U}_{n}^{(2)}\big)}
{\overline{{\cal U}_{n}^{(2)}\left(0\right)}}
= \frac{\sqrt{{\cal S}_{\xi\xi}(j)\lambda^{2}
+ {\cal S}_{\xi\delta}(j)\lambda + {\cal S}_{\delta\delta}(j)}}
{2(\lambda+1)}.
\label{eq:sigmasglgr}
\end{equation}
Figure~\ref{fig:sigma} shows this quantity plotted as a function of $j$ for both GOE and GUE.  Notice that (i) the rms is of the same order as the average, and (ii) its magnitude weakly depends on $j$.  To show the magnitude of the fluctuations in the correction relative to its average value, we plot two additional curves in both Fig.~\ref{fig:ave1} and Fig.~\ref{fig:ave3}, namely $\delta u \pm \sigma_{\delta u}$.  We find that at the realistic value $j \!=\! 0.3$, the correction to the even peak spacing is \textit{negative} for a small fraction of the quantum dots in the ensemble.

\section{Conclusions}
\label{sec:conclusions}

In this paper we studied corrections to the spacings between Coulomb blockade conductance peaks due to finite dot-lead tunneling couplings.  We considered both GOE and GUE random matrix ensembles of 2D quantum dots with the electron-electron interactions being described by the CEI model. We assumed $T \!\ll\! \Delta \!\ll\! E_{C}$. The $S \!=\! 0$, $\frac{1}{2}$, and $1$ spin states of the QD were accounted for, thus limiting the applicability of our results to $J_{\rm s} \!<\! 0.5\Delta$.  

The ensemble-averaged correction in even valleys is given in Eq.~(\ref{eq:even2}).  The average correction decreases monotonically (always staying positive, however) as the exchange interaction constant $J_{\rm s}$ increases (Figs.~\ref{fig:ave1}-\ref{fig:goe}).  The behavior found is very robust with respect to the choice of RMT ensemble or change in charging energy, mean level spacing, or temperature.  Our results obtained in second-order perturbation theory in the tunneling Hamiltonian are somewhat similar to the zeroth-order results\cite{Usa01,Usa02} in that the exchange interaction reduces even-odd asymmetry of the spacings between peaks.  While the average correction to the even spacing is positive, that to the odd peak spacing is negative and of equal magnitude.

The fluctuations of the correction to the spacing between Coulomb blockade peaks mainly come from the mesoscopic fluctuations of the wave functions and energy level spacing $x$ in the QD.  The rms fluctuation of this correction is given by Eqs.~(\ref{eq:varianceref})-(\ref{eq:xi}).  It is of the same order as the average value of the correction (Figs.~\ref{fig:ave1} and \ref{fig:ave3}) and weakly depends on $J_{\rm s}$ (Fig.~\ref{fig:sigma}).  Therefore, for a small subset of ensemble realizations, the correction to the peak spacing at the realistic value of $j = 0.3$ is of the opposite sign. The rms fluctuation of the correction for an odd valley is the same as that for an even one.

We are aware of two experiments directly relevant to the results here.
First, in the experiment by Chang and co-workers,\cite{Jeo} the corrections to the even and odd peak spacings due to finite dot-lead tunnel couplings were measured.  It was found that the even (odd) peak spacing increases (decreases) as the tunnel couplings are increased.  This is in qualitative agreement with the theory; see Eq.~(\ref{eq:even2}).  The magnitude of the effect was measured at different values of the gas parameter $r_{s}$ (and, hence $J_{\rm s}$) as well.  Unfortunately, because the effect is small and the experimentalists did not focus on this issue, one cannot from this work draw a quantitative conclusion about the behavior of the correction to the peak spacing as a function of $J_{\rm s}$.

Second, in the experiment by Maurer and co-workers,\cite{Mau99} the fluctuations in the spacing between Coulomb blockade peaks were measured as a function of the dot-lead couplings with $g_{L} \!=\! g_{R}$.  Therefore, only the symmetric part [Eq.~(\ref{eq:sigmasym7})] would contribute to the total variance.  Reference~\onlinecite{Mau99} reported results for two dots: a small one with area 0.3\,$\mu$m$^{2}$ and a large one with area 1\,$\mu$m$^{2}$.  From the area (excluding a depletion width of about 70 nm), we estimate that the large (small) dot contains about 500 (100) electrons. Measurements on the large QD found larger fluctuations upon increasing the dot-lead tunnel coupling, in qualitative agreement with the theory. Though the temperature was larger than the mean level spacing in the large QD whereas our theory is developed for $T \!\ll\! \Delta$, the theory gives about the correct magnitude for the peak spacing fluctuations. It is inconclusive whether the data is in better agreement with the CI or CEI model as the fluctuations are roughly the same (Fig.~\ref{fig:sigma}) in both. In the small QD, there is an anomaly for the strongest coupling in the experiment -- the fluctuations suddenly decrease. In addition, the experimental fluctuations are one order of magnitude larger than the theoretical estimate [Eqs.~(\ref{eq:sigmasglgr}) and (\ref{eq:un20})]. The reason for this discrepancy is not clear at this time. Possible contributing factors include scrambling of the electron spectrum as the charge state of the dot changes, or the role of the fluctuations when the dot is isolated (i.e., fluctuations in ${\cal U}^{(0)}$). In order to assess quantitatively the role of dot-lead coupling in the Coulomb blockade, further experiments are needed.

\begin{acknowledgments}
We are grateful to A.~M.~Chang, A.~M.~Finkelstein, A.~Kaminski, C.~M.~Marcus, K.~A.~Matveev, L.~I.~Glazman, and G.~Usaj for stimulating discussions.  This work was supported in part by NSF Grant No. DMR-0103003.  
\end{acknowledgments}


\begin{thebibliography}{50}
\bibitem{KMq}
More precisely, the conductance peak has its maximum at the gate voltage corresponding to the maximum of the total amplitude of the tunneling through many energy levels in a QD: ${\cal A}=\sum_{k\sigma}{\cal A}_{k\sigma}$, see Ref.~\onlinecite{Kam00}.  Therefore, in addition to the dominant resonant term that gives the charge degeneracy point result for the peak maximum [Eq.~(\ref{eq:energy})], the total tunneling amplitude includes elastic cotunneling terms (we assume that $T\ll\Delta$) with random phases.  However, one can show (see Ref.\,\onlinecite{Kam00} for details) that these cotunneling terms give a negligible fluctuating correction to the position of the peak maximum with $\mbox{mean}\big(\delta{\cal N}_{n-\frac{1}{2}}\big)\approx 0$ and $\mbox{rms}\big(\delta{\cal N}_{n-\frac{1}{2}}\big) \approx\left(g_L+g_R\right)^{3}/\left(4\pi\right)^{3}\xi$.  The coefficient in the last equation corresponds to the CI model.
\bibitem{Border}
One may think that an additional term due to shift in the border $x = 2j$ between singlet and triplet states should be present in this equation as well.  However, one can show that the contribution of this term to $\overline{{\cal U}_{n}^{(2)}}$ (as well as to its rms) is small.
\end{thebibliography}
\end{document}